\documentclass[5p, times]{elsarticle}

\usepackage{hyperref}
\usepackage{lineno}
\usepackage{xcolor}
\newcommand{\galight}{\textsc{galight}}
\newcommand{\sersic}{S\'ersic}
\newcommand{\lenstronomy}{\textsc{lenstronomy}}

\newcommand\aj{{AJ }}
\newcommand\apj{{ApJ }}
\newcommand\pasp{{PASP }}
\newcommand\aap{{A\&A }}
\newcommand\mnras{{MNRAS }}

\usepackage{color, listings, courier, relsize}
\usepackage{amsmath}
\usepackage{amsfonts}
\usepackage{amssymb}
\usepackage[htt]{hyphenat}

\definecolor{Code}{rgb}{0,0,0}
\definecolor{Decorators}{rgb}{0.5,0.5,0.5}
\definecolor{Numbers}{rgb}{0.5,0,0}
\definecolor{MatchingBrackets}{rgb}{0.25,0.5,0.5}
\definecolor{Keywords}{rgb}{0,0,1}
\definecolor{self}{rgb}{0,0,0}
\definecolor{Strings}{rgb}{0,0.63,0}
\definecolor{Comments}{rgb}{0,0.63,1}
\definecolor{Backquotes}{rgb}{0,0,0}
\definecolor{Classname}{rgb}{0,0,0}
\definecolor{FunctionName}{rgb}{0,0,0}
\definecolor{Operators}{rgb}{0,0,0}
\definecolor{Background}{rgb}{0.98,0.98,0.98}
\definecolor{Booleans}{rgb}{0.572,0,0.572}
\definecolor{BuiltinFunction}{rgb}{0.572,0,0.572}
\definecolor{BuiltinConstant}{rgb}{0.572,0,0.572}
\definecolor{Asterisk}{rgb}{0.670,0,1}

\lstdefinelanguage{Python}{
    	numbers=left,
    	numberstyle=\footnotesize,
    	numbersep=7pt,
    	xleftmargin=1.26em,
    	framextopmargin=2em,
    	framexbottommargin=2em,
    	showspaces=false,
    	showtabs=false,
    	showstringspaces=false,
    	frame=l,
    	tabsize=4,
    	stepnumber=1,
	basicstyle=\ttfamily\small,
    	backgroundcolor=\color{Background},
    	breaklines=True,
    	postbreak=\mbox{\textcolor{red}{$\hookrightarrow$}\space},
	commentstyle=\color{green}\ttfamily,
    	stringstyle=\ttfamily\color{Strings},
    	morecomment=[s][\color{Strings}]{'}{'}, 
    	stringstyle=\ttfamily\color{Comments},
    	morecomment=[s][\color{Comments}]{\#}{\#}, 			
	stringstyle=\ttfamily\color{Strings},
	morekeywords={import,from,class,def,for,while,if,is,in,elif,else,not,and,or,print,break,continue,return,access,as,except,exec,finally,global,import,lambda,pass,print,raise,try,assert},
    	keywordstyle={\color{Keywords}\bfseries}, 
    	morekeywords={[2]True,False,None},
    	keywordstyle={[2]\color{BuiltinConstant}\slshape},
	otherkeywords={[2]*},
	keywordstyle={[2]\color{Asterisk}},
	emph={self},
	emphstyle={\color{self}\slshape}	
}

\lstdefinelanguage{bash}{
    numbers=none,
    numberstyle=\footnotesize,
    numbersep=7pt,
    xleftmargin=1.26em,
    framextopmargin=2em,
    framexbottommargin=2em,
    showspaces=false,
    showtabs=false,
    showstringspaces=false,
    frame=none,
    tabsize=4,
    stepnumber=1,
    basicstyle=\ttfamily\small,
    backgroundcolor=\color{Background},
    breaklines=True,
    postbreak=\mbox{\textcolor{red}{$\hookrightarrow$}\space},
}

\modulolinenumbers[5]

\journal{Journal of \LaTeX\ Templates}

\begin{document}

\begin{frontmatter}

\title{Galaxy shapes of Light (GaLight): a 2D modeling of galaxy images}
\tnotetext[mytitlenote]{For \galight\ version 0.1.0}





\author[ipmu,corres]{Xuheng Ding}
\author[kipac,slac]{Simon Birrer}
\author[ucla]{Tommaso Treu}
\author[ipmu,UoT]{John D. Silverman}
\address[ipmu]{Kavli Institute for the Physics and Mathematics of the Universe, The University of Tokyo, Kashiwa, Japan 277-8583 (Kavli IPMU, WPI)}
\address[kipac]{Kavli Institute for Particle Astrophysics and Cosmology and Department of Physics, Stanford University, Stanford, CA 94305, USA}
\address[slac]{SLAC National Accelerator Laboratory, Menlo Park, CA, 94025}
\address[ucla]{Department of Physics and Astronomy, University of California, Los Angeles, CA, 90095-1547, USA}
\address[UoT]{Department of Astronomy, School of Science, The University of Tokyo, 7-3-1 Hongo, Bunkyo, Tokyo 113-0033, Japan}
\address[corres]{email: xuheng.ding@ipmu.edu}


\begin{abstract}
\galight\ is a Python-based open-source package that can be used to perform two-dimensional model fitting of optical and near-infrared images to characterize the light distribution of galaxies with components including a disk, bulge, bar, and quasar. The decomposition of stellar components has been demonstrated in published studies of inactive galaxies and quasar host galaxies observed by the Hubble Space Telescope and Subaru's Hyper Suprime-Cam. \galight\ utilizes the image modeling capabilities of \lenstronomy\ while redesigning the user interface for the analysis of large samples of extragalactic sources. 
The package is user-friendly with some automatic features such as determining the cutout size of the modeling frame, searching for PSF-stars in field-of-view, estimating the noise map of the data, identifying all the objects to set the initial model and associated parameters to fit them simultaneously. These features minimize the manpower and allow the automatic fitting tasks.  The software is distributed under the MIT license.
The source code, installation guidelines, and example notebooks code can be found at \url{https://galight.readthedocs.io/en/latest/}. 
\end{abstract}


\end{frontmatter}


\section{Introduction} \label{sec:intro}
Accurate surface photometry and morphology of galaxies are key to understanding the evolution of galaxies. Over the past few decades, the light distribution of galaxies has been successfully studied using analytic functions to estimate their size, luminosity, concentration, orientation and symmetry. Of all tested analytic models, the \sersic\ model function~\citep{sersic} is the most successful and commonly used, which provides an accurate description of the surface brightness of galaxies, ranging from exponential disks (i.e., late-type galaxies) to de Vaucouleurs's profiles \cite{deVaucouleurs1948} for describing early-type galaxies. 
The success of these analytic functions further allow us to analyze the light of a unique type of object, i.e., active galactic nuclei (AGN), which has been recognized to play an important role in the evolution galaxies and the growth of its central supermassive black hole (SMBH). While a central AGN can be bright, even above that of its host galaxy, we can still infer the photometry and morphology of the underline host galaxy based on the use of analytic models and deblending techniques.

With current and upcoming large imaging surveys, the sample of galaxies with quality imaging is growing rapidly and will continue to do so with future facilities. Thus, there is a great demand to process such large numbers of galaxy images in an automated manner with efficiency and accuracy.

Here, we present a first public release of \galight, an open source \textsc{python} software package to model galaxy images including those hosting AGN. \galight~uses routines available through \lenstronomy~\citep{Birrer2018, Birrer2021} that are further described below. \galight~adds a number of features to process large numbers of images from survey data. This tool has been used to model the galaxy images from Hubble Space Telescope (HST)~\citep{Ding2020, Bennert2021} and Subaru Hyper Suprime-Cam (HSC)~\citep{Tang2021, Li2021}, including a sample of $\sim$1.5 million galaxies and $\sim$5 thousand quasars profiting from the automatic features.

While other packages, such as \textsc{Galfit}~\citep{Peng2002} and \textsc{GIMD2D} \citep{Simard1998}, have provided the community much needed capabilities for analyzing galaxy images, there are benefits available with \lenstronomy\ which we aim to fully exploit. For example, the parameter estimation is based on a forward modeling approach with semi-linear inversion of the amplitude components using machinery that includes a particle swarm optimizer (PSO) \citep{Kennedy1995, Birrer2015} and Markov Chain Monte Carlo (MCMC) error analysis by \textsc{emcee}~\citep{emcee}. The ability to account for uncertainties on the point spread function (PSF) is also an important feature. Equally important, \lenstronomy\ -- an \textsc{astropy}-affiliated \textsc{python} tool for gravitational lensing image simulations and analyses -- is an open source software package that can be easily installed and adapted to the user's modeling needs due to its flexibility.

While functions provided by \lenstronomy\ are suitable to model the light distribution of a galaxy, some additional work is needed to define the models, set the initial parameters and priors, implement masks for nearby objects, and assess specific requirements for each image cutout. As described here, the software package \galight\ can set all the necessary parameters and perform automatic modeling with much needed efficiency, particularly for the analysis of large samples. Below, we introduce the basic functions of \galight\ and demonstrate its capabilities.

The paper is organized as follows. In Section~\ref{sec:overview} and \ref{sec:fit-pro}, we present a brief overview of the package which is structured in three major classes. In Section~\ref{sec:modules}, we provide detailed information with some example code and figures. We highlight features in the Appendix and a summary is given in Section~\ref{sec:con}. We refer the reader to a number of studies that have made use of \galight~\citep{Ding2020, Bennert2021, Tang2021, Li2021}.

\section{Package overview} \label{sec:overview}
\galight\ is open source python-based~\citep{van1995python} software with distribution granted through the MIT open source software license. The software relies on the \textsc{python} standard library and the open-source libraries including \textsc{numpy}~\citep{numpy}, \textsc{scipy}~\citep{scipy}, \textsc{astropy}~\citep{astropyI, astropyII}, \textsc{matplotlib}~\citep{matplotlib} and \textsc{photutils}~\citep{photutils}. In particular, the package \lenstronomy~\citep{Birrer2018, Birrer2021} is the work horse that performs the fitting process. \galight\ is compatible with both \textsc{python}~2 and \textsc{python}~3. However, the latter version is recommended\footnote{To run \galight\ on a \textsc{python}~2 environment, the \lenstronomy\ in version 1.3.0 should be installed.}. 

The public release version of \galight\ is on Pypi\footnote{\url{https://pypi.org/}} and can be installed using \texttt{pip}:
\begin{lstlisting}[language=bash]
$ pip install galight --user
\end{lstlisting}
We also distribute \galight\ on \texttt{GitHub}\footnote{\url{https://github.com/dartoon/galight}}, while more stable versions released through Pypi. Documentation is available through sphinx\footnote{\url{www.sphinx-doc.org}} and released on \texttt{ReadtheDocs}\footnote{\url{https://galight.readthedocs.io/}}. To fully demonstrate the usage of this package, we provide several \texttt{Jupyter}\footnote{\url{http://jupyter.org}} notebooks with extended features. 

\galight\ helps the user to execute model fitting with ease based on, but not limited to, automated features as listed as follows:

\begin{enumerate}
\item Determining a suitable cutout size for the image frame. The cutout should have a minimal size that balances the need to reduce computing time and capture the extend of low surface brightness emission. The cutout will also keep neighboring galaxies away from the edge of the frame. This enables these galaxies to be properly modeled which may have outer light profiles that overlap with the central target of interest.

\item 
The PSF model is key for image (de-)convolution and accounting for an unresolved point source component (i.e., AGN and small unresolved components). The common practice is to manually select a star in the field of view (FOV) for characterizing the PSF\footnote{Another practice for large imaging surveys (Dark Energy Survey (DES) or in the upcoming LSST) is to generate an interpolated PSF model, e.g., PSF Extractor \href{https://www.astromatic.net/software/psfex/}{(\texttt{psfEX}}).}. \galight\ will search for all the PSF-stars in the FOV, report their full width at half maximum (FWHM), and allow the user to choose those suitable for determining the empirical PSF model.

\item If needed, a noise map can be generated based on the exposure time and the background noise level measured from empty regions.

\item Neighboring sources can be simultaneously modeled or easily be masked out.

\item By default, all the objects in the modeling frame will be detected by the \texttt{photutils.detect\_sources}. Based on the detection results, the parameter settings for all these sources will be assigned automatically.

\item Output data products are generated for full assessment of the goodness-of-fit with the ability to share across different computing platforms.
\end{enumerate}

In practice, \galight\ is able to perform the fitting tasking without manual supervision after the target position and the modeling materials are provided.

\section{Model-fitting procedure}
\label{sec:fit-pro}

\galight\ uses three steps to achieve a successful modeling of image data. Here, we provide a brief overview of each step as defined by one \textsc{python} class and provide further details in Section~\ref{sec:modules}.

\begin{enumerate}
\item \texttt{DataProcess:} The data products are prepared including the cutout of the science image, noise map, PSF model, and nearby object masks (optional). The image pixel scale and magnitude zero point should also be provided to \galight\ to calculate physical quantities (i.e., target magnitudes and galaxy size in arcsec).

\item \texttt{FittingSpecify:} The model components (e.g., single \sersic\ profile, disk$+$bulge, AGN, etc.) and associated parameters should be defined by this class. For example, the initial values and priors of the parameters are assigned. Priors can be given as either a range for each parameter with uniform or weighted sampling, or a single fixed value. 

\item \texttt{FittingProcess:} The model fitting is performed, and the results are presented in this step. The user can determine the method of parameter inference (e.g., minimization and MCMC). Once the fitting has completed, this class provides the modeling output for diagnostic and demonstration purposes. 
\end{enumerate}

\subsection{Analytical model and fitting strategies}
Through \lenstronomy, the model components are defined and the fitting is performed. The imported features include the definition of the light profile and the approach to sampling the parameter space.
\\
\\
{\bf Light profiles}: All extended light profiles defined by \lenstronomy\ are imported and available. These include, but not limited to, \sersic~\citep{sersic}, Chameleon~\citep{Dutton2021}, ellipsoid, 2D Gaussian, pseudo Jaffe~\citep{Jaffe}, Moffat~\citep{Moffat}, and power-law functions. We refer the reader to the \lenstronomy\footnote{\url{https://lenstronomy.readthedocs.io/en/latest/lenstronomy.LightModel.html}} documentation for further details.

By default, the \sersic\ profile~\citep{sersic} is adopted to describe the light of extended object with parameterization as follows: 
\begin{eqnarray}
   \label{eq:sersic}
   &I(R) = A \exp\left[-k\left(\left(\frac{R}{R_{\mathrm{eff}}}\right)^{1/n}-1\right)\right] ,\\
   &R(x,y,q) = \sqrt{x^2+y^2/q^2},
\end{eqnarray}
where $A$ is the amplitude\footnote{Note that $A$ is also the surface brightness at $R_{\mathrm{eff}}$.} and $q$ denotes the axis ratio. 
The S\'ersic index $n$ controls the shape of the radial surface brightness profile, a larger $n$ corresponds to a steeper inner profile and a highly extended outer wing.  $k$ is a constant which solely depends on $n$ to ensure that the isophote at $R=R_{\mathrm{eff}}$ encloses half of the total light~\citep{Ciotti1999}. Note that this definition of \sersic\ profile is the same as the one adopted by \textsc{Galfit} in which the $R_{\mathrm{eff}}$ is the semi-major-axis half-light radius. The other definition of elliptical light profile with a format as $R(x,y,q) = \sqrt{qx^2+y^2/q}$ is also available in \lenstronomy\ after version 1.9.0, hence \galight. Both formats will be available to the user\footnote{The definition of semi-major-axis is designed to be used by default, expect that the \lenstronomy\ version 1.9.0 is installed through \texttt{pip}. For this particular \lenstronomy\ version, the user needs to change the corresponding configure \href{https://github.com/sibirrer/lenstronomy/blob/main/lenstronomy/Conf/conf_default.yaml}{{\sc file}} to change the half-light radius definition. Eventually,  \texttt{FittingSpecify.sersic\_major\_axis} records which definition is adopted for the fitting.}; however, note that the $R_{\mathrm{eff}}$ will be different by a factor of $\sqrt{q}$ between these two definitions.

We note that the light profile for spheroidal galaxies can usually be fitted with a single \sersic\ profile and a free index $n$. In cases where more components are required, a composite light model can be implemented, such as a spheroid plus disk component (\sersic\ $n$ = 1) and even a bar (\sersic\ $n$ = 0.5).

When modeling the point-source flux component, 
such as an AGN component or star, a scaled PSF model will be adopted to characterize the unresolved emission. The amplitude and the position of the point source will be considered as free parameters. As demonstrated in next section, a PSF model can be directly input to \galight. Alternatively, \galight\ can search for suitable stars over the full image to generate an empirical PSF model. To account for an imperfect model PSF, one can also provide an uncertainty on the PSF which will be considered when calculating the likelihood and best-fit model parameters. Taking the PSF model uncertainty into account can be important, especially when estimating the uncertainty of the fitting parameters. This is likely an important feature of \galight\ which
is not a feature directly implemented in the currently most commonly used modeling codes. 
To note, the PSF model is also used for the convolution of the total emission in the forward-modeling.

{\bf Fitting algorithm}: To obtain the inference of the best-fit parameters, a likelihood is calculated by \lenstronomy\ using the \texttt{lenstronomy.ImSim} module while sampling of the parameter space is performed by the \texttt{lenstronomy.Sampling} module.
Briefly, a likelihood estimator is implemented to determine the values of the posterior distribution based on the data and model $p(d_{\rm data}|d_{\rm model})$ (i.e., the probability of the data given a model) using all the counted pixels in the full cutout which can be presented as follows:
\begin{eqnarray}
   \label{eq:likelihood}
    \log p(d_{\text{data}}|d_{\text{model}}) = \sum_i \frac{(d_{\text{data,i}} - d_{\text{model,i}})^2}{2\sigma_{\text{i}}^2} + C.
\end{eqnarray}

The errors (i.e., $\sigma_{\text{i}}$) are assigned based on the noise map. For fitting, the noise map can be directly input to \galight. If not provided, the tool can calculate the total noise map according to the following two terms:

\begin{equation}
\label{eq:errormap}
    \sigma_{i}^2 = \sigma_{\text{bkgd}}^2 + |d_{\text{data,i}}/(t_{\rm exp} \times G_{\rm eff})|,
\end{equation}

\noindent i.e., a Gaussian background term (i.e., $\sigma_{\text{bkg}}$), and a Poisson term, i.e., $\sqrt{|d_{\text{data,i}}/(t_{\rm exp} \times  G_{\rm eff})|}$ 
, where $t_{\rm exp}$ is the exposure time (in seconds) of the observational data (given in counts per second) and $G_{\rm eff}$ is the effective gain values.
By default, the \sersic\ surface brightness amplitude (i.e., the parameter $A$ in Equation~\ref{eq:sersic}) is linear parameters that is solved directly during the likelihood estimation. This modeling strategy saves a type of non-linear parameter and would lead to better and faster convergence; even more, it can be generalized to more components and basis sets.


\section{Learning to use \galight\ through examples} \label{sec:modules}

We provide detailed descriptions of the routines and three core classes. In this paper, we give brief show-case examples and further demonstrations can be found online through Jupyter notebooks\footnote{\url{https://github.com/dartoon/galight_notebooks}}.
\subsection{DataProcess}

The following example in next page is a demonstration of the steps needed to prepare the class \texttt{DataProcess}. We provide example code (see next page) and highlight the relevant lines being discussed.
\begin{lstlisting}[language=Python,float=*]
import astropy.io.fits as pyfits
#Load the data, noise map, PSF and header information based on astropy#
dataFile = pyfits.open('../example_data.fits')
fov_image = dataFile[1].data
header = dataFile[1].header
noiseFile = pyfits.open('../example_noise.fits')
noise_map = noiseFile[1].data

#Prepare for the fitting#
from galight.data_process import DataProcess
pos_RA, pos_Dec = 500, 500   #Assuming the target is at pixel position 500, 500#
zp = 27.0 #Assuming the zeropoint of the detection is 27.0#
rm_bkglight = True #The background light will be modeled and removed as a 2D profile by photutils.background.SExtractorBackground() .#

#Input the information to DataProcess.#
data_process = DataProcess(fov_image = fov_image, target_pos = [pos_RA, pos_Dec], fov_noise_map = noise_map, header = header, rm_bkglight = rm_bkglight, zp = zp)

#Preparing fitting ingredients, including cutout image, noise map and masks (if needed).#
data_process.generate_target_materials(radius=65, create_mask = True, nsigma=2.8, exp_sz= 1.2, npixels = 15)
#Automatically find all the PSF-stars and select the best one to use for fitting.#
data_process.find_PSF(radius = 30, user_option = True) 
#Alternatively, the users can load a PSF model manually through:#
#data_process.PSF_list = [PSF]#

#Check out if all the fitting ingredients are ready. If not, an error message will appear.#
data_process.checkout()
\end{lstlisting}

\galight\ retrieves the pixel scale from the image header (line~16), or this is set by the user manually \texttt{data\_process.deltaPix}. The location of the target can be given as either a pixel position or with `WCS' information as RA and DEC (line~11). The function \texttt{generate\_target\_materials} (line~19) extracts a square region centered on the target and creates an image stamp having a frame size of $(2\times$\texttt{radius}$+1)$. If the \texttt{radius} value is assigned as `\texttt{None}', \galight\ will determine a minimum value, ranging from 30 to 70 pixels, that avoids having nearby objects not fully covered by the cutout image. Of course, for nearby galaxies, the upper boundary might not be sufficient and the user should select their own value.

\begin{figure*}
\centering
\begin{tabular}{c c}
\hspace*{-1cm}
{\includegraphics[height=0.3\textwidth]{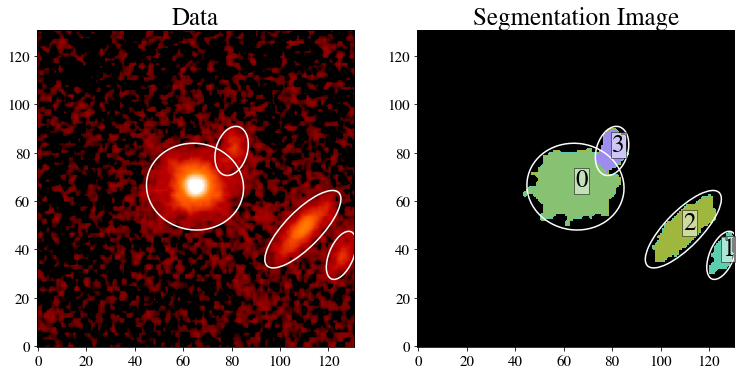}}&
{\includegraphics[height=0.3\textwidth]{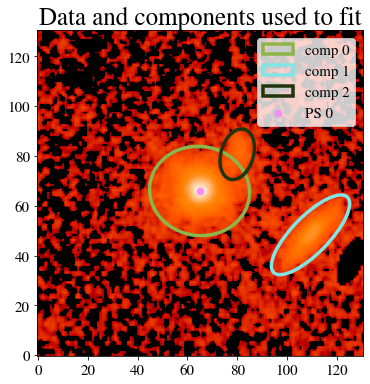}}\\
\end{tabular}
\caption{Example target with HST cutout images. {\it Left~panel}: Detected objects in the cutout frame are highlighted by white elliptical apertures. 
{\it middle~panel}: Each object is labeled with individual numbers to help the user to select those to be masked. {\it right~panel}: After user input, the final settings show how this image will be modeled. All apertures indicate the initial settings to fit the extended sources. `PS0' marks the center of the fitted point source of interest. Here, object \#1 has been masked out thus filled in black in the bottom right panel.
}
\label{fig:fitting_sets}
\end{figure*}

Objects in the frame are detected 
as part of the \texttt{generate\_target\_materials} function,
and characterized by a set of elliptical {\it apertures} as shown in Figure~\ref{fig:fitting_sets} (left and middle panels). The user is given the option to either mask any object or model them simultaneously in the fitting procedure (see notebook entitled 
\href{https://github.com/dartoon/galight_notebooks/blob/v0.1.0/galight_HST_QSO.ipynb/}{~`\texttt{galight\_HST\_QSO.ipynb}'}
for more details).
As shown in the figure, the object~\#0 is our fitting target which is assigned by aperture~\#0. The other nearby objects are also assigned by the corresponding apertures. These apertures are saved in `\texttt{data\_process.apertures}', and by default they reveal how the data will be fitted as a sets of elliptical \sersic\ profiles; the properties of these apertures (including positions, size, ellipticity and orientation) will be used to assign the initial parameters of the \sersic\ profiles to model.
One can manually change the properties of apertures in this step to modify the initials that will be used in the fitting. The user can also add the apertures manually  to increase the number of \sersic\ profiles that to be fitted. For example, if we aim to fitting our targeting galaxy as disk$+$bulge, we can creat a smaller aperture (i.e., a bulge component) and add it to the position at the aperture~\#0.  The \textsc{python} notebook
\href{https://github.com/dartoon/galight_notebooks/blob/v0.1.0/galight_HSC_galaxy.ipynb/}{~`\texttt{galight\_HSC\_galaxy.ipynb}'} demonstrates how this setting can be achieved. The prior settings for the \sersic\ profiles are introduced in next section.
In addition, these apertures can be used to assign the masks to block the pixels for any object in the fitting. In this example, we mask out the object~\#1 in the fitting as shown in the Figure.

For this example, the noise level of the data is known and directly loaded (see line 16, the \texttt{noise\_map}, in the example code). If this information is not provided, the \texttt{generate\_target\_materials} function will automatically measure the background noise level from empty regions and combine this with the Poisson noise component. We demonstrate the estimation of the noise map in~\ref{app_a}. 

A model PSF is required for the convolution and, if needed, as an individual component to represent an unresolved point source. If the PSF model is available with an image, it can be input to the \texttt{DataProcess()} class directly (see line 23 in the example code). Alternatively, a PSF can be constructed based on stars in the field-of-view of the science image. We provide a functionality to search for PSF stars across the entire image in \texttt{data\_process.find\_PSF()}. The user can then select the preferred objects based on the sharpness of their light profile or for any other reason. A demonstration of how to search for suitable objects to generate an empirical model PSF is provided in~\ref{app_b}. Online notebooks (
\href{https://github.com/dartoon/galight_notebooks/blob/v0.1.0/find_PSF_in_FOV.ipynb/}{~`\texttt{find\_PSF\_in\_FOV.ipynb}'}
and 
\href{https://github.com/dartoon/galight_notebooks/blob/v0.1.0/galight_HST_QSO.ipynb/}{~`\texttt{galight\_HST\_QSO.ipynb}'})
demonstrate this feature.
To assign an error on the PSF, one can define an array type that has the same dimensions as the PSF (see next section). 


\subsection{FittingSpecify}\label{sec:fitspec}

In this class, the light model and parameters are set. Since the aim of \galight\ is to automatize the process, there are very few steps required for the user. An example is as follows.
\begin{lstlisting}[language=Python]
#Produce the class and params to make the QSO decomposition.#
from galight.fitting_specify import FittingSpecify

#Pass the data_process to FittingSpecify#
fit_sepc = FittingSpecify(data_process)

#Prepare the fitting sequence, point_source_num defines the number of point source (i.e., AGN) that to fit.#
fit_sepc.prepare_fitting_seq(point_source_num = 1)

#Plot the initial settings for fittings. #
fit_sepc.plot_fitting_sets()

#Build up all the settings and get prepared to pass to the next class.#
fit_sepc.build_fitting_seq()
\end{lstlisting}
    
The apertures determined by \texttt{DataProcess()} in previous step are applied and the initial \sersic\ parameters are set as starting points for the fitting. 
The main settings of the fit routine are controlled by \texttt{prepare\_fitting\_seq()} which take arguments that can be set as follows:

\begin{enumerate}
\item \texttt{point\_source\_num} sets the number of point sources needed for the fit (default = 0). This should be set to one for a quasar or a star. By default, the software will automatically find the source with brightest peak in the cutout frame and use the peak position as the initial position to perform the fit. 

\item \texttt{fix\_center\_list} sets the priors on the position so that an AGN and its host galaxy are fixed to the same central position.

\item \texttt{fix\_n\_list} can be assigned to fix the \sersic\ index value and \texttt{fix\_Re\_list} can be assigned to fix the \sersic\ R$_{\rm eff}$ value.

\item \texttt{psf\_error\_map} defines the PSF error map with the same dimensions as the input PSF. The values of the error map are added at the fitted point source position to the final noise map as:\\
$\sigma_{i}^2 
= \sigma_{i}^2$ +  \texttt{psf\_error\_map}$_j * ($\texttt{point\_source\_flux}$)^2$.
\end{enumerate}
\noindent Examples of these setting can be found in the online notebooks. Detailed information is given in the online \href{https://galight.readthedocs.io/en/latest/galight.fitting_specify.html#module-galight.fitting_specify}{documentation} through \texttt{ReadtheDocs}.

The \texttt{prepare\_fitting\_seq()} is designed to help the user quickly build up the requisites for the fitting using the templates embedded in the function. After this step, the user can make modifications to these settings to add complexity and make the fitting more sophisticated. For example, the \texttt{FittingSpecify.kwargs\_params} is a dictionary that saves all the parameters settings for the \sersic\ model and the point sources including the initials, sampling steps, fixed value, lower and upper limit. The user can change the corresponding values therein to modify the lower and upper limits for any parameters so that these parameters will be limited in the desired range and sampled with a flat prior. Additional priors can also be added to the likelihood. For example, defining the disk size to be larger than the bulge size or any arbitrary prior on the parameters (e.g., a Gaussian prior), can be assigned through the \texttt{FittingSpecify.kwargs\_likelihood}. These are demonstrated in \href{https://github.com/dartoon/galight_notebooks/blob/v0.1.0/galight_HSC_galaxy.ipynb/}{~`\texttt{galight\_HSC\_galaxy.ipynb}'}.
Note that all this settings are in the format that defined by \lenstronomy; the user should follow its guidance to achieve more sophisticated fittings.

Finally, the user can use \texttt{plot\_fitting\_sets()} to display how the target and nearby sources will be modeled as shown here in the right panel of Figure~\ref{fig:fitting_sets}.

\subsection{FittingProcess}
\label{sec:fit_pro}

The fitting routine is executed after \texttt{FittingSpecify} has been passed to \texttt{FittingProcess}:
\begin{lstlisting}[language=Python]
#Setting the fitting method and run:#
from galight.fitting_process import FittingProcess
#Pass fit_sepc to FittingProcess:#
fit_run = FittingProcess(fit_sepc, savename = 'savename')
#Setting the fitting approach and run. The fitting process will be save as 'savename.pkl'#
fit_run.run(algorithm_list = ['PSO', 'MCMC'], setting_list = [None, None]))
\end{lstlisting}

By default, the procedure first performs the minimization of the parameters using the 
particle swarm optimizer (~\citep[PSO,][]{Kennedy1995, Birrer2015}) algorithm and then passes the inferred minimized parameters to the MCMC routine (e.g., \textsc{emcee}~\citep{emcee}) to estimate their best-fit values and uncertainties. In the \texttt{FittingProcess.run()}, i.e., line 6 in the example, the user can re-define the \texttt{algorithm\_list} to increase the times of minimization or skip the MCMC through the arguments. Moreover, the sampling parameters such as the fitting particles and the number of iterations can be defined through \texttt{setting\_list}.

\begin{figure*}
\centering
\includegraphics[width=1\textwidth]{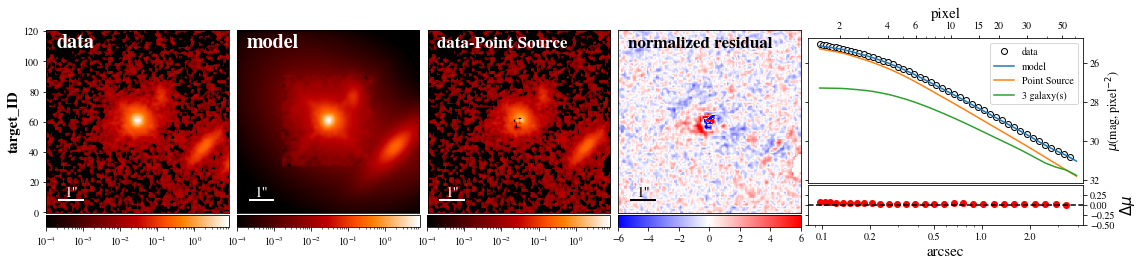}
\includegraphics[width=0.85\textwidth]{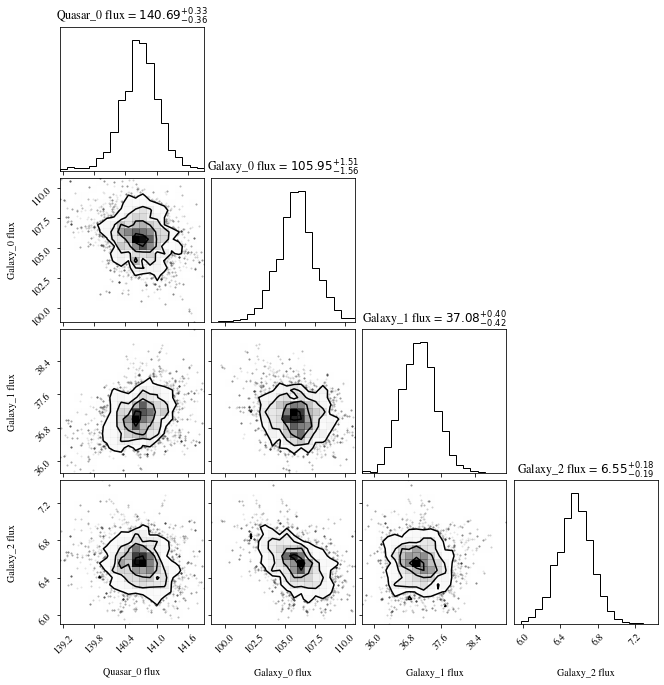}
\caption{$top~panels$: Example modeling results of the image data shown in Figure 1. The panels from the left to right are: (1) observed science data, (2) best-fit model. The clear box around the galaxy reveals the region of the PSF size that used to model the AGN. (3) data - point-source model (i.e., host galaxy only) (4) residuals divided by variance (i.e., reduced $\chi^2$ map), (5) surface brightness profiles in 1D (top) and the residual (bottom). $Bottom~panels$: the MCMC inference of the likelihood distribution of the fitted parameters are translated to flux values (i.e., the total pixel counts) for each fitted component. The labels for each component correspond to those in the right panel of Figure~\ref{fig:fitting_sets}. Note that the flux values are not the parameters that fitted directly.
}
\label{fig:fitting_result}
\end{figure*}

Once parameter estimation has finished, the trace of the PSO and MCMC fitting chains will be plotted to help the user diagnose the convergence of the fitting routine. 
\begin{lstlisting}[language=Python]
#Fitting has done. Now, plot the fitting results (as shown in Figure 2, top panel):#
fit_run.plot_final_qso_fit()
#Calculate the physical properties, such as magnitude, for each component:#
fit_run.translate_result()
#Plot the MCMC inference of the flux for each component (as Figure 2, bottom panel):#
fit_run.plot_flux_corner()
\end{lstlisting}
The MCMC parameter inferences are translated to fluxes values and displayed in the bottom panel of Figure~\ref{fig:fitting_result} with the 
posteriors on the flux of the individual components.

\section{Conclusion}~\label{sec:con}
We present an open-source python-based tool --\galight-- to decompose galaxies into their various components. The software is distributed under the MIT license. Taking advantage of many of the functionalities of \lenstronomy, the modeling is composed of three steps, separated into individual classes. We briefly introduce the functions of each class and present the example results in Section~\ref{sec:modules}. \galight\ provides additional functions to setup the fitting routine and complete the modeling in an automated manner. Our aim is to minimize intervention by the user so that large data sets can be processed seamlessly.

The demonstrations and other features mentioned in this paper are included in example notebooks in~\url{https://github.com/dartoon/galight_notebooks} as listed here.

\begin{enumerate}
\item `\texttt{galight\_HST\_QSO.ipynb}': Modeling example of an HST observed quasar image. In this example, the PSF is reconstructed by a PSF-star in the FOV and the noise map is calculated using \galight\ based on the approach introduced in~\ref{app_a}.

\item `\texttt{galight\_HSC\_QSO.ipynb}': Modeling example of an HSC imaged quasar, in which the PSF and noise map are supplied to \galight.

\item `\texttt{galight\_HSC\_dualAGN.ipynb}': Modeling example of an HSC imaged double AGN candidate.

\item `\texttt{galight\_HSC\_galaxy.ipynb}': Modeling example of an HSC imaged galaxy. As demonstration, the galaxy is modeled as composite model with disk and bulge components.
\end{enumerate}
\noindent These online resources will be updated periodically.
\\
\\

{\bf Citing the code}:
This manuscript is posted on arXiv as a user manual, which will be constantly updated in the future with more upcoming features. We ask the users of \galight\ please make reference to
\href{https://doi.org/10.3847/1538-4357/ab5b90}{Ding et al., ApJ, 2020, 888, 37} and \lenstronomy~\citep{Birrer2021} when use it.

\section{Acknowledgements}
This software is supported by World Premier International Research Center Initiative (WPI), MEXT, Japan.

%

\vspace{5mm}



\appendix

\section{Noise map estimation}
\label{app_a}

\begin{figure*}
\centering
\includegraphics[width=0.7\textwidth]{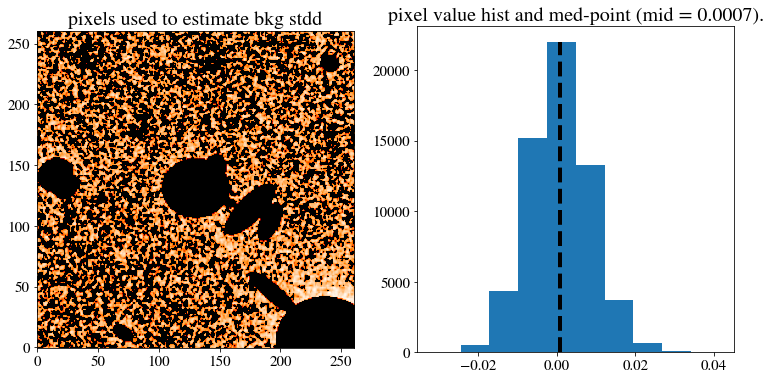}
\caption{{\it left}: Measuring the background noise level with  \galight. The emission from the objects in the field are detected and masked so that the sky pixels can be identified. {\it right}: Histogram of the number counts from sky pixels. The black dashed line is the median value. Since the background light has been removed in our example, the position of the median value is located at zero.
}
\label{fig:est_bkg}
\end{figure*}

If user does not provide an error map, \galight\ will  calculate the noise level for each pixel which is composed of a Gaussian background noise and Poisson noise components, (see Equation~\ref{eq:errormap}). 
First, \galight\ detects all objects and then masks them to identify the empty (sky) regions to measure the average signal. We demonstrate an example of this measurement in Figure~\ref{fig:est_bkg}. A Gaussian formalization is used to measure the standard derivation. Since the background light has been subtracted by setting \texttt{rm\_bkglight = True} in \texttt{DataProcess()}, the median of the distribution should close to zero, which can be easily checked (Figure~\ref{fig:est_bkg}; right panel). The Poisson noise is estimated based on the pixel number counts and the exposure time, where exposure time can be input as either a single value or a array map. Alternatively, the user can measure the Gaussian background themselves and directly input the error map to \galight.

\section{Search for PSF-stars in the image FOV}\label{app_b}
\galight\ can search for available PSF-stars in the FOV of the image using function \texttt{DataProcess.findPSF}. The algorithm is designed so that it looks for the local maximum pixel with prominent features (like PSF), achieved by the function \texttt{scipy.ndimage.filters}. During this search process, the code measures the FWHM values of each PSF candidates and discards those with extended profiles. Moreover, PSF-stars that are too faint or bright, compared with the target will be removed by default.

\begin{figure*}
\centering
\includegraphics[width=0.85\textwidth]{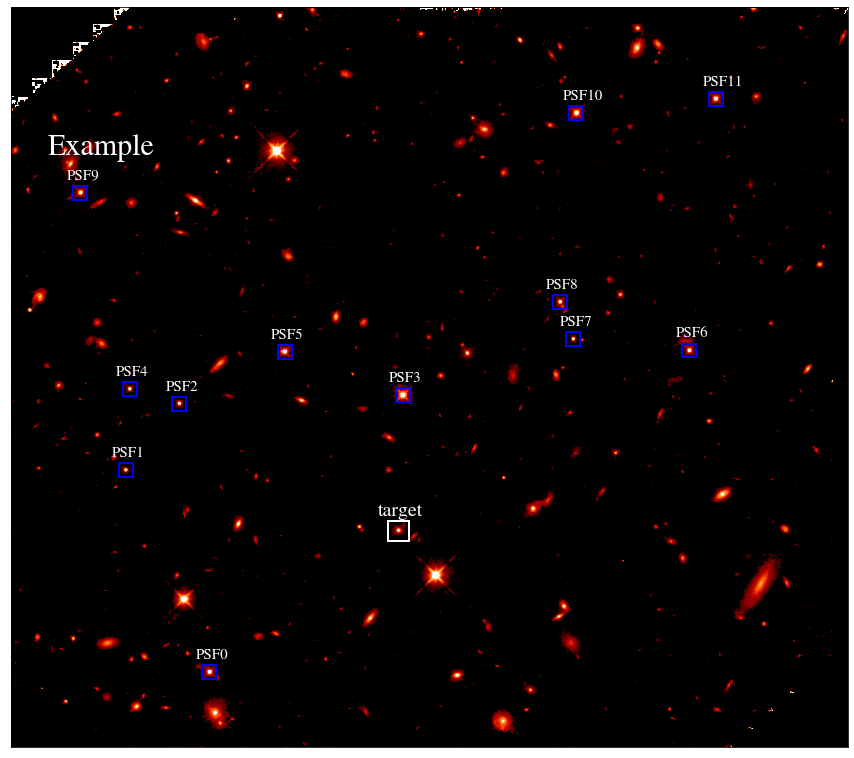}
\caption{Identifying suitable PSF stars in a HST/WFC3 field. Candidates found  by  \galight\ are marked by small blue boxes and checked as to whether they are not too bright or too faint.}
\label{fig:findPSF}
\end{figure*}

\begin{figure*}
\centering
\includegraphics[width=0.6\textwidth]{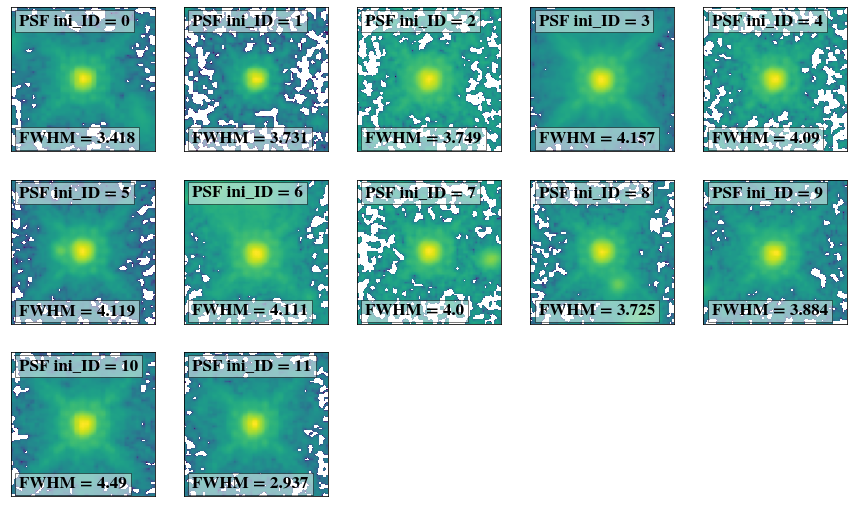}
\caption{Zoom-in images of the PSF model candidates. During this process, the FWHM (in units of pixel scales) of each PSF is measured and reported to help the user select the best PSFs for their application.
}
\label{fig:PSF_library}
\end{figure*}

\begin{figure*}
\centering
\includegraphics[width=0.5\textwidth]{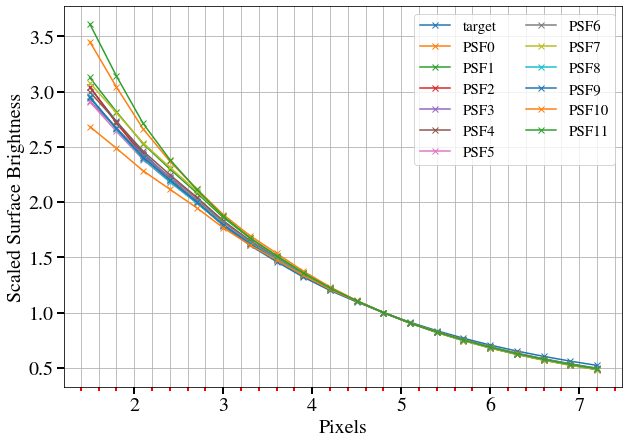}
\caption{The 1D surface brightness profile of each PSF model candidates compared to the targets.
}
\label{fig:PSF_profile}
\end{figure*}

We demonstrate the search of PSF stars in the FOV. The FWHM and 1D profile is also presented to the user to aid in the selection of the best PSF to use for the fitting (see Figures~\ref{fig:findPSF}, \ref{fig:PSF_library} and \ref{fig:PSF_profile}).
By default, the PSF with the sharpest profile (i.e., the one with smallest FWHM value) will be selected. 

\end{document}